\magnification=\magstep 1
\baselineskip=16pt
\overfullrule=0pt
%\normallineskip=8pt
%\vsize=23 true cm
%\hsize=15 true cm
%\voffset=-1.0 true cm
%\font\bigfont=cmr10 scaled\magstep1
%\footline={\hss\tenrm\folio\hss}
\newcount\fignumber
\fignumber=0
\def\fig#1#2{\advance\fignumber by1
 \midinsert \vskip#1truecm \hsize14truecm
 \baselineskip=15pt \noindent
 { {\bf Figure \the\fignumber} #2}\endinsert}

\def\ref#1{$^{[#1]}$}
\def\sqr#1#2{{\vcenter{\vbox{\hrule height.#2pt
   \hbox{\vrule width.#2pt height#1pt \kern#1pt
   \vrule width.#2pt}\hrule height.#2pt}}}}

\def\ref#1{[#1]}
%------------------------ paper ---------------------
\bigskip\bigskip
\centerline{\bf PHASE TRANSITIONS IN GRANULAR PACKINGS}
\bigskip\bigskip\bigskip\bigskip\smallskip
\centerline{\bf A. Coniglio$^{a,b}$ and H.J.~Herrmann$^a$}
\medskip
\centerline{$^a$P.M.M.H. (U.R.A. 857), E.S.P.C.I. Paris}
\centerline{10 rue Vauquelin, 75231 Paris, France}
\centerline{$^b$Dipartimento di Scienze Fisische, Universit\`a di Napoli}
\centerline{INFN and INFM Sezione di Napoli}
\centerline{Mostra D'Oltremare Pad.19, 80125 Napoli, Italy}
\bigskip\bigskip\bigskip\vskip 3.5truecm
\noindent{\bf Abstract}

{\noindent
We describe the contact network of granular packings by a frustrated 
lattice gas that contains steric frustration as esential ingredient.
Two transitions are identified, a spin glass transition at the
onset of Reynolds dilatancy and at lower densities a percolation
transition. We describe the correlation functions
that give rise to the singularities and propose some 
dynamical experiments .}
\vskip 1.5truecm\noindent
PACS: 05.70, 46.10 \hfill
\vskip 3.0truecm
Let us consider a packing of granular material, like
sand or powder, at rest. Depending on how the packing
has been prepared its density and its rheological properties can
be different. If the sand is poured into a recipient its
volumetric density is relatively low and it can be
moved like a fluid. A stick can be inserted into it
and removed again easily since the structure of the grains 
is rather loose. If the recipient is then 
shaken at low amplitude the level of
sand visibly decreases evidencing an increase of the density.
Below a certain level a stick that was put inside before
shaking cannot be removed anymore unless a large
force is applied\ref{1}. This effect evidences
the existence of a critical density, described by
Reynolds\ref{2} on 1885, above which the granular structure cannot be 
sheared anymore. The grains are interlocked so densely that
large deformations become impossible unless the grains are separated
by some amount, i.e. the global density is decreased again.
On the wet beach this is the effect which produces the dry sand around
the pressure zone of our foot step.
This Reynolds dilatancy has been carefully studied experimentally 
and numerically by 
soil mechanicists\ref{3} and has been included into the
equations of motion of plastic deformation\ref{4}.

%The density at the onset of Reynolds dilatancy seems to separate
%two phases: Above it shear is resisted and below it not.
%Can one describe this onset as an equilibrium phase transition
%in the thermodynamic sense? 
Apparently there are many possible configurations of grain positions that
lead to a static packing. They can be characterized by their
density. Only a subset of these configurations is sufficiently dense
to give rise to mechanical locking. To describe these different types
of packings
Sam Edwards and collaborators\ref{5,6} have formulated a thermodynamic
description of a powder which replaced the energy by
the volume and the temperature by the compactivity $\kappa = 
\partial V/\partial S$. We will describe the ensemble
of static packings by adapting to granular media a statistical
mechanical model which combines spin glass and lattice gas properties
and which appears to describe also the phenomenology of the glass forming
liquids\ref{7-11}.

The guiding idea of our approach is to consider each static packing as
a local minimum in some free energy functional. The rough landscape in
configurational space induces slow relaxation phenomena and may
exhibit non-ergodic behaviour. The classical mechanism leading to such
a complex structure of local minima is based on the concept of
frustration. In a granular packing this frustration arises because the
grains having different shapes interlock and
%and are therefore intrinsically unable to
%form close-packed configurations at high densities. Frustration also
%arises in a kinematic sense: 
due to steric hindrance cannot
move locally except through large-scale cooperative rearrangements of
many particles that usually change the volume.

In this paper we elaborate on a lattice gas model~\ref{7-11} containing
frustration as an essential ingredient. This model has been called
``frustrated percolation'' (FP)~\ref{8} because it can be viewed as
percolation~\ref{12} in a frustrated medium. Both static and dynamic   
properties of the model exhibit complex behaviour with features that
are common to granular packings, structural glasses and spin
glasses. This may suggest that, just as the Ising  model provided
insight into the physics of critical phenomena near second-order
phase transitions, the frustrated lattice gas model presented here may
help to understand the Reynolds transition through the spin glass
transition.  

Let us consider a square lattice tilted in such a way that the axis of
the lattice forms $45^{\circ}$ with the horizontal. Gravity points
vertically down. We assign to each edge of the lattice an interaction
$\epsilon_{ij}=-1$ with probability $\pi$ and $\epsilon_{ij}=+1$ with
probability $1-\pi$. We assume for now that this assignement is fixed
in space, i.e. that the disorder is quenched. We also fix $\pi=1/2$. A
loop is considered frustrated if the product of the sign along the
loop is negative. Next we introduce particles on the vertices of the
upper part of the lattice and let them move down randomly with the
constraint that frustrated loops cannot be fully occupied, i.e. that
at least one site must be missing in each frustrated loop. The
particle stops moving when either all vertices below are occupied or
if any move downwards would violate the constraint. So, the constraint
induces defects or holes into the system, since a frustrated loop must
necessarily contain at least one empty site. Therefore when no more
particles can be poured into the system from above the system will
have a density $\rho_m$ less than unity, i.e. the density when every
site would be occupied by one particle.

One can now allow the particles to diffuse by exchanging their
positions with a nearest-neighbor hole such that no frustrated loop
becomes completely occupied. Due to gravity we consider the
probability $p_1$ to move upwards to be less than the probability
$p_2$ to move downwards. We let the system evolve for a time $t$ after
which the probability $p_1$ to move upwards is put equal to zero and
the system is left to relax until the particles do not move
anymore. We expect that the final density will be lower than before
the exchange of positions. The smaller
the ratio $x=p_1/p_2$, and the larger the time $t$ the lower we expect
the final density.  At low particle densities frustration does not
significantly inhibit diffusion. At high densities, however, 
it can happen that a
particle can only move through the system by long-range collective
rearrangements of many particles. On experimental or numerical time
scales this particle will then be frozen within a cluster of particles
which interlock each other. Such a frozen cluster is depicted
schematically in Fig.~1. On very long time scales frozen clusters can
dissolve, which is why they have also been called ``quasi-frozen'' in
the literature~\ref{7}. The extremely long life time of frozen
clusters is responsible for the slow relaxation phenomena at high densities.
%Physically the process described
%above would for instance correspond to applying a vibration to the
%system, with fixed sufficiently high frequency (so that only single particle
%movements are obtained, too low frequency should induce the movement of many
%particles at the same time) and amplitude A. Large amplitudes
%corresponds to large $x$ values.

\fig{4}{Schematic representation of (a) a spanning cluster of
particles (b) a spanning cluster of frozen particles. Filled circles
are frozen particles and empty circles are particles that can diffuse.}

The case $p_1=p_2$ and $t=\infty$ is identical to ``site-frustrated
percolation'' as introduced before~\ref{8}. In fact also a
``bond-frustrated percolation'' has been introduced~\ref{7,9} in which the
bonds instead of the sites are occupied and which has the same
critical properties as the site model. The static properties of
bond-frustrated percolation were studied via a renormalization group
calculation~\ref{9} on a hierarchical lattice, using a Hamiltonian
formalism which will be introduced in the Appendix . The results have been
confirmed by Monte Carlo simulations on the square
lattice~\ref{10}. Two critical points were found, one at a density
$\rho_p$ which is in the universality class of the ferromagnetic
1/2-state Potts model, while the second critical point at $\rho_R$ is
in the same universality class as the $\pm J$~Ising spin glass
transition. Each critical point is characterized by a diverging
length, associated with the quantities
$$p_{ij} = p_{ij}^+ + p_{ij}^-\eqno(1)$$ and
$$g_{ij} = p_{ij}^+ - p_{ij}^-\ \ .\eqno(2)$$ Here
$p_{ij}^+(p_{ij}^-)$ is the probability that on one hand the sites $i$
and $j$ are connected by at least one path of bonds and on the other
hand the product over all signs (or ``phase'')
$\epsilon_{kl}$ along the path connecting $i$ and $j$ is +1(-1).
In the granular packing a path between
$i$ and $j$ corresponds to a chain of touching grains that connect
grains $i$ and $j$. The length $\xi_p$ associated with
the pair connectedness function $\langle p_{ij} \rangle$ 
diverges at the percolation density $\rho_p$, while the length 
$\xi$ associated with $\langle g_{ij}^2 \rangle$
diverges at the higher density $\rho_R$, where 
$\langle .. \rangle$ is the average
over all possible interaction configurations $\{ \epsilon_{ij} \}$.
The dynamic properties of site-frustrated percolation with
$p_1=p_2$ have also 
been studied numerically on a two dimensional square 
lattice\ref{13}. It was found that the system below the percolation
transition,
develops a dynamical behaviour in the density-density autocorrelation
function, characterized by a two step relaxation function typical of
glass forming liquids. Finally as the temperature approaches the high
density 
critical point $\rho_R$ the system freezes with the diffusion
coefficient going to zero.

The rich scenario described above, including the transition at
$\rho_R$ occurs due to the freezing of paths, i.e. the
interference of 
different phases. If the particles are free to move the 
cancelation on average of paths of different phases
would make $g_{ij}$ very small.
The only way to maintain the asymmetry in the phase
of the paths, needed to have a large $g_{ij}$ is by developping frozen
paths. The length $\xi$ is a measure of the size of the clusters of
frozen sites (see fig.~1).  At the critical density $\rho_R$
the length $\xi$ diverges, 
i.e.  the different paths connecting two grains $i$ and $j$
being arbitrarily far apart can block each others motion freezing the
entire system mechanically. This is precisely the phenomenology of the
Reynolds transition as described in the introduction.  It seems
therefore plausible that while $\rho_p$ corresponds to the lowest
density for which a static packing can be achieved, the density
$\rho_R$ corresponds to the onset of Reynolds dilatancy. Since FP
close to $\rho_R$ exhibits the same static properties of the spin glass
transition and dynamical properties closely related to glass
forming liquids, we expect that granular media near the Reynolds
transition should have the static properties of a spin
glass and the dynamical properties ressembling that of a glass forming
liquid.

In the rest of this paper we want to
elucidate more deeply these properties and propose experimental
consequences of this equivalence to granular packings. In the Appendix
we establish the relation to the
Hamiltonian description of spin glasses.

Spin glasses have been studied extensively~\ref{14} and are known to
have a ``spin glass transition'' at temperature $T_{SG}$ in three
dimensions. This transition is characterized by a diverging
correlation length associated with the square of the spin-spin
correlation function, and consequently exhibits a strong divergence in
the non-linear susceptibility. The specific heat instead exhibits a
weak singularity characterized by a negative exponent $\alpha$, very
difficult to detect experimentally or numerically. The FP also
exhibits a diverging length associated to $g_{ij}$. However the strong
divergence associated to the sum over the sites of $g_{ij}$, is not
accessible by an experimental probe. Instead the compressibility
associated to density fluctuations has the same weak behaviour as
the spin glass specific heat.

To pursue the analogy between granular systems and spin glasses it
seems useful to consider also the thermal excitations of a packing. A
granular temperature $T$ has been introduced by Ogawa~\ref{15,16} as
being proportional to the kinetic energy of the particles. This can be
implemented experimentally by putting the sand into a box which is
vibrated at constant frequency $\omega$. The box should not be
too large to ensure that the vibration energy is homogeneously
distributed inside the system. Varying the amplitude $A$ of
the vibration one can control the temperature via $T \propto
A^2$. On can, starting with a given $A_0$, decrease the amplitude
linearly with time $t$, like $A(t) = A_0 - r t$ with a rate $r$.
It has been shown numerically~\ref{13} that in the 
site-frustrated percolation model, 
the volume per particle $v$ as function of the temperature exhibits a plateau
below a temperature $T_g(\dot T)$ which is the lower the lower the cooling rate
$\dot T$. Consequently the final density when the temperature 
$T\rightarrow 0$ is higher for smaller cooling rates.
This effect is also found in real
glass forming liquids but not reproducible within the simple spin glass model.
Since in our analogy $r \sim \dot T$ we predict that
in the above described experiment of lowering the
vibration amplitude with time one should obtain different
final densities for different rates $r$. In fact $r$ would
be a good control parameter for the Reynolds transition.
It would be interesting to experimentally verify this prediction.

Another parameter to control the final density could
be the  amplitude $A$ itself in the case in which
the vibration is suddenly stopped and 
the system very rapidly
falls into a static configuration. Since the quenching takes a time
that is of the order of the period $2\pi \over \omega$ of the
vibration the cooling rate $\dot T$ monotonically increases with
$A$ so that then the control parameter steering the Reynolds
transition would be $A$. A more detailed investigation of this
dependence would also be interesting.  

It is experimentally not easy to measure dynamical
auto-correlation functions or the local response to
a mechanical perturbation in particularly in three
dimensions. More accessible is the macroscopic relaxation
of a granular packing. The analogy to glass forming liquids
would suggest that at high densities
the restructuring after a perturbation should be slower
than logarithmic. We think that doing Reynold's classic
bag experiment~\ref{2} in which a bag filled with sand and water
is deformed and the water level is seen to go down, might
be a candidate to measure this slow relaxation, namely
by measuring how fast the level increases again with time.
For such an experiment, however, the elastic properties of
the bag will have to be carefully controlled.

By describing a granular packing through a lattice model in which
particles must fulfill non-local, quenched constraints we found that 
the configurational ensemble is equivalent to that of frustrated
percolation. This later model is known to exhibit a percolation
and a spin glass transition and its dynamical properties
ressemble those of glass forming liquids. It seems plausible to
identify the spin glass transition to the onset of Reynolds dilatancy.
In reality, however, the steric constraints in granular systems
are not quenched in space but determined by the actual positions of the 
grains. Considering annealed disorder instead would smoothen the
percolation transition but leave the spin glass transition
essentially unaltered~\ref{7}. Configurations that are obtained
starting from a given granular packing (for instance through shaking)
do, however, have memory of the original disorder, so that the
experimental reality is neither completely quenched nor completely
annealed. Further investigations in this direction are under way.
\bigskip \noindent{\bf Appendix:}
\medskip The general Hamiltonian formalism needed to describe the
site-bond frustrated percolation problem is given by:
$$-H = J \sum_{ij} [(\epsilon_{ij}S_iS_j +1)\delta_{ \sigma_i
\sigma_j}-2] n_i n_j + \mu \sum_i n_i \eqno(3)$$ where $n_i$ is the
occupancy of site $i$ ($n_i = 1$ if site occupied and $n_i = 0$ if
site empty), $S_i$ a spin variable on an occupied site, having the
states $\pm 1$.  $\sigma_i = 1....s$ are Potts variables which assume
$s$ states, $\mu$ is the chemical potential for a particle and $J$ is
related to the chemical potential for a bond $\mu_b$ through
$e^{\mu_b/kT} = (1-e^{-2J/kT})/e^{-2J/kT}$.  In the limit $s
\rightarrow 1/2$ the Hamiltonian (3) reproduces site-bond frustrated
percolation.  In this model the configurations are made of sites
connected by bonds. The bonds cannot occupy entirely a frustrated
closed loop.  As a particular case in the limit $\mu \rightarrow \infty$ 
($\mu_b \rightarrow \infty$) and $s
\rightarrow 1/2$ we obtain frustrated bond (site) percolation while in
the limit $\mu \rightarrow \infty$ and $s \rightarrow 1$, we obtain
the $\pm$~J Ising spin glass model.

\bigskip\bigskip\bigskip
\centerline {\bf References}
\bigskip\medskip
\item{1.} This effect was recently studied by V. Horv\'ath and
I. Janosi, preprint HLRZ 67/95, in preparation
\item{2.} Reynolds O., Phil. Mag. S. {\bf 20} (1885) 469
\item{3.} Bashir Y.M. and Goddard J.D., J. Rheol. {\bf 35} (1991) 849
\item{4.} Vermeer P.A. and de Borst, Heron {\bf 29} (1984) 1
\item{5.} Edwards S. F., J. Stat. Phys. {\bf 62} (1991) 889
\item{6.} Mehta A. and  Edwards S. F., Physica A {\bf 157} (1989) 1091
\item{7.} Coniglio A., Il Nuovo Cimento {\bf 16D} (1994) 1027
\item{8.} Coniglio A., in {\it Correlation and Connectivity},
eds. H.E. Stanley and N. Ostrowsky (Kluwer Acad. Publ., 1990),
Coniglio A., di Liberto F., Monroy G. and Perrugi F., Phys. Rev. B
{\bf 44} (1991) 12605 
\item{9.} Pezzella U. and Coniglio A., to be published
\item{10.} de Candia A., Cataudella V. and Coniglio A., in preparation
\item{11.} Glotzer S.G. and Coniglio A., Computational Material
Science {\bf 215} (1995) 1.
\item{12.} Stauffer D. and Aharony A., {\it Introduction to
percolation theory}, 2nd edition (Taylor and Francis, London, 1994)
\item{13.} S. Scarpetta et al., in preparation 
\item{14.} For a review see Binder K. and Young P., Adv. in Physics
{\bf 41} (1992) 547
\item{15.} Ogawa S., Proc. of US-Japan Symp. on Continuum Mechanics
and Statistical Approaches to the Mechanics of Granular Media,
eds. S.C. Cowin and M. Satake (Gakujutsu Bunken Fukyu-kai, 1978), p.208
\item{16.} H.J. Herrmann, J. Physique II {\bf 3} (1993) 427
\end